\documentclass[a4paper,twoside]{JHEP}
\usepackage{epsfig}
\def\sezpi{{1\over 16\pi^2}}
\def\trdpi{{1\over 32\pi^2}}
\newcommand{\be}{\begin{equation}}
\newcommand{\en}{\end{equation}}
\newcommand{\bea}{\begin{eqnarray}}
\newcommand{\ena}{\end{eqnarray}}
\newcommand{\lbl}[1]{\label{eq:#1}}
\newcommand{\rf}[1]{(\ref{eq:#1})}
\def\d{\partial}
\def\D{\nabla}
\def\dm{g_m}
\def\ddm{h_m}
\def\em{e_m}
\def\fm{f_m}
\newcommand{\lapprox}{%
\mathrel{%
\setbox0=\hbox{$<$}\raise0.6ex\copy0\kern-\wd0\lower0.65ex\hbox{$\sim$}}}
\newcommand{\gapprox}{%
\mathrel{%
\setbox0=\hbox{$>$}\raise0.6ex\copy0\kern-\wd0\lower0.65ex\hbox{$\sim$}}}
\setbox0=\hbox{$\scriptstyle\circ$}
\newcommand{\dmo}%
{\mathrel{\setbox1=\hbox{$g_m$}\copy1\kern-0.8\wd1\raise1.1\ht1\copy0
\kern0.6\wd1}}
\newcommand{\ddmo}%
{\mathrel{\setbox1=\hbox{$h_m$}\copy1\kern-0.8\wd1\raise1.1\ht1\copy0
\kern0.6\wd1}}
\newcommand{\emo}%
{\mathrel{\setbox1=\hbox{$e_m$}\copy1\kern-0.8\wd1\raise1.1\ht1\copy0
\kern0.6\wd1}}
\newcommand{\fmo}%
{\mathrel{\setbox1=\hbox{$f_m$}\copy1\kern-0.8\wd1\raise1.1\ht1\copy0
\kern0.6\wd1}}
\preprint{MIT-CTP-2977\\IPNO/DR-0010}

\title
{Flavour Stability of the Chiral Vacuum and Scalar Meson Dynamics}

\author
{ Bachir Moussallam\thanks{Permanent address: Groupe de 
Physique Th\'eorique, IPN Bat100, Universit\'e Paris-Sud, 91406 Orsay,
France} \\
Center for Theoretical Physics\\
Massachusetts Institute of Technology\\
Cambridge, MA 02139\\}

\abstract{
Previous work relating the flavour variation of the chiral
order parameters $F_\pi$, $\langle\bar u u\rangle$ and S-wave
scattering data, based on chiral
sum rules and chiral perturbation theory at order $p^4$, is extended 
to include $O(p^6)$ corrections. The finding of a significant decrease
of these order parameters, particularly $\langle\bar u u\rangle$, 
with the number of flavours 
increasing from $N_F=2$ to $N_F=3$ is confirmed,
modulo an assumption on the convergence of the chiral expansion. 
The connection between scalar resonance physics and the
phase structure of the chiral vacuum is also 
illustrated on the basis of the linear sigma-model.
We allow for a very general symmetry breaking sector compatible with softness.
The result depends strongly on the input scalar meson masses, in particular,
on the presence, or not, of a light sigma.}


\begin{document}
\section{Introduction}

Chiral perturbation theory for $N_F=3$ light flavours
at next-to-leading order was shown by Gasser
and Leutwyler\cite{gl85} to involve ten independent 
coupling constants $L^r_i(\mu)$ (referred to, sometimes, as LEC's). 
In ref.\cite{gl85} all ten couplings were determined from  low-energy 
data except two, $L_4$ and $L_6$. According to the OZI rule or,
alternatively, large $N_c$ considerations these couplings are expected to be
suppressed relative to the other ones\cite{gl85}. 
Physically, $L_4$ and $L_6$ control
how observables of the pion, like $M_\pi$ or 
chiral $SU(2)$ order parameters 
$(F_\pi)_{SU(2)}$, $\langle \bar u u\rangle_{SU(2)}$ 
vary if the mass of
the strange quark mass $m_s$ varies. If OZI suppression holds these quantities
are expected to be essentially insensitive to variations of $m_s$
and the value of the chiral order parameters
$F_\pi$,  $\langle \bar u u\rangle$ in the $SU(2)$ chiral limit
or in the $SU(3)$ chiral limit (i.e. $m_s=0$) ought 
to be very nearly the same.
In nature, of course, the values of the light quark masses are fixed but
they can be made to vary in unquenched lattice simulations of QCD.  

There are  reasons to suspect that the couplings $L_4$ and $L_6$ are,
in fact, far from being suppressed.  One reason is suggested by two recent
unquenched lattice simulations which have investigated the phase structure
of QCD-like theories with $N_F$ equal-mass flavours when the value of 
$N_F$ is varied. It was found in ref.\cite{iwasaki} that for $N_F>6$ a phase 
with no confinement and no spontaneous chiral symmetry breaking prevails, at
any value of the QCD coupling constant. In ref.\cite{mawhinney} a very strong
decrease of the chiral order parameters was observed upon
varying $N_F$ from
$N_F=2$ to $N_F=4$. The values of the coupling constants $L_i$ encode
informations on the physics of the massive states in QCD\cite{egpr}. 
In particular, $L_4$, $L_6$ are linked to the scalar
resonances. It is notorious that the OZI or large $N_c$ rules seem to
fail in this sector. In fact,  is not clear at present exactly which
of the scalar resonances form the lowest lying nonet (see e.g. the
reviews in the pdg\cite{pdg}). In other words, the fact that $L_4$, $L_6$
may be unsuppressed, the failure of the OZI rule in the scalar sector
are connected and these features may be related in an interesting way to the
phase structure of QCD, in particular to the fact that a phase transition
could occur for a value of $N_F$ not exceedingly larger than $N_F=3$.
An interpretation of these features in terms of a paramagnetic effect
of the quark loops is discussed in ref.\cite{descotes}

These topics have started to be 
investigated in a previous paper\cite{L6}. The couplings
$L_4,\ L_6$ were related to the scalar resonances via the scalar form-factors
of the pion and the Kaon.
These form-factors can be reconstructed 
directly from experimental data on S-wave scattering, modulo a few
plausible hypothesis\cite{dgl}. The coupling $L_4$ is related to the derivative
of the strange form-factor of the pion at the origin and, for $L_6$, 
a chiral sum rule can be derived in the terms of the correlator $\Pi_6$
of the strange scalar current $\bar s s$ and the non-strange current
$\bar u u+\bar d d$. In the present paper, we investigate the $O(m_s)$
corrections to the results of ref.\cite{L6}, which could
be sizable. For this purpose, one needs to use CHPT at order $p^6$. 
Renormalizability of the theory at this order was recently proved\cite{bce1},
and  the set of independent chiral lagrangian terms classified\cite{bce2}.
In particular, we will use the expansions of $M_\pi$ and $F_\pi$ which
were derived in refs.\cite{kambor} and \cite{amoros} and we have computed
the $O(p^6)$ expansion of the correlator $\Pi_6$. The result of this effort,
at first sight, will turn out to be mitigated: 
as new coupling constants appear
in the formulas, there is a loss in predictivity over the leading order 
calculation. As discussed in sec.2,  conclusions may nevertheless be
drawn provided one makes 
some assumptions on the convergence of the chiral expansion.

In order to gain further insight, we have also investigated a different, more
model dependent approach in sec.3. Following the idea of 
ref.\cite{egpr} one starts from a
lagrangian for the scalar resonances and the LEC's are generated by 
integrating out the resonances. If we attempt to determine some $O(p^6)$
LEC's in this way, from the most general lagrangian, we find that there 
are many undetermined resonance parameters 
which get involved, like three resonance
couplings, couplings of one resonance to two scalar sources etc... In sec.3, 
we discuss the more specific dynamics generated by the linear 
sigma-model\cite{lsm}.
This model is still attracting interest\cite{vautheri}\cite{randrup} 
in connection  with the phenomenon of disordered
chiral condensate\cite{cdd} which could be formed in heavy ion collisions. 
In the
symmetry breaking sector, we will consider a lagrangian more general
than previously done, still compatible with the criterion of softness
(restricting ourselves to first order symmetry breaking). 
The model can accommodate exactly, in principle, a given set
of scalar nonet masses and one obtains the $O(p^4)$ and $O(p^6)$ LEC's as 
definite predictions. The question at this point is which scalar mesons
are to be included in the nonet? In particular, does one have to include
a light $\sigma$ and a light $\kappa$ meson? It has long been known that
an S-matrix pole can be identified in low-energy $\pi\pi$ 
scattering\cite{basdevant} with a very large 
imaginary part. A similar structure was argued to exist
also in $\pi K$ scattering\cite{beveren}\cite{fariborz1}, 
but it is unclear whether such
poles should be interpreted as physical scalar resonances. For our
purposes, we will leave this question open and consider two different
possibilities for the scalar nonet. As we will see, in the framework
of the sigma-model,  these will correspond
to rather different behaviour of the chiral vacuum. 

\section{$L_4$, $L_6$ beyond $O(p^4)$}

Our estimates of $L_4$ and $L_6$ are based, essentially, on a method
to extract  the scalar form-factors
of the pion and the kaon using experimental $\pi\pi-K\bar K$ scattering
data, proposed in ref.\cite{dgl}. The form-factors get determined up to a
normalization factor, for which one uses CHPT. In order to determine $L_4$ one
equals the experimental determination of $G'_\pi(0)$, the derivative
of the strange scalar form-factor of the pion and the chiral expansion of this 
quantity. In ref.\cite{L6} this matching was performed at leading order 
in the chiral expansion, i.e. $G'_\pi(0)$ was expanded up to $O(p^4)$ and
for the normalization condition we used CHPT at order $p^2$. In a similar
way, for $L_6$ we use the scalar form factors to express the spectral function
of the correlator $\Pi_6(s)$ (see \rf{Pi6def} below) and $L_6$ is obtained
by matching $\Pi_6(0)$ evaluated from experimental data and its chiral
expansion. Below, we discuss the chiral corrections to these results, which
involve two parts a) one must use the chiral expansions of $G'_\pi(0)$ and
$\Pi_6(0)$ up to  $O(p^6)$ and b) we must use CHPT at $O(p^4)$
instead of $O(p^2)$ in the normalization conditions. 

Let us designate by $F_\pi(s)$ and $G_\pi(s)$ the
non-strange and strange scalar form factors of the pion, and by
$F_K(s)$ and $G_K(s)$ the analogous ones for the Kaon. We will
use the following normalizations,
\bea\lbl{norm}
&F_\pi={\displaystyle1\over\displaystyle B_0}
\sqrt{\displaystyle3\over\displaystyle2}\langle0\vert \bar u u
+\bar d d\vert\pi^0\pi^0\rangle\quad
&G_\pi={1\over B_0}\sqrt{3\over2}\langle0\vert \bar s s
\vert\pi^0\pi^0\rangle\nonumber\\
&F_K={\displaystyle1\over\displaystyle B_0}
\sqrt{\displaystyle2}\langle0\vert \bar u u
+\bar d d\vert K^+ K^-\rangle\quad
&G_K={1\over B_0}\sqrt{2}\langle0\vert \bar s s
\vert K^+ K^-\rangle\ .
\ena
The method of ref.\cite{dgl} consists in solving numerically
a set of Muskhelishvili-Omn\`es  coupled-channel
equations using experimentally determined $\pi\pi-K\bar K$
S-wave scattering T-matrix elements 
as input(in practice we made use of the parametrisations
of ref.\cite{au} and ref.\cite{kll}). Strictly speaking, these equations hold
under the assumption of exact two-channel unitarity up to $s=\infty$
while in practice, two-channel unitarity is a good approximation up to
the $\eta\eta$ threshold. The resulting form factors are expected to
be reliable in a finite energy range, which we will assume to extend
up to 1 GeV. One also chooses appropriate boundary conditions for the 
T-matrix at $s=\infty$ which insure existence of a solution with a 
minimal number of free parameters: a solution vector $u_1(s),\ u_2(s)$
will be determined in the entire energy range once the values at one point,
say at $s=0$, are given.
Extension to more channels
and the stability of this scheme were discussed in ref.\cite{L6}.
In practice, one first constructs numerically 
two independent solutions of the equation set: $u_i(s)$ and $v_i(s)$, $i=1,2$,
normalized at the origin such that $u_1(0)=1$, $u_2(0)=0$ and  
$v_1(0)=0$, $v_2(0)=1$. The form factors are then given as
\be\lbl{Fu1u2}
{F_\pi(s)\choose F_K(s)}=F_\pi(0)
{u_1(s)\choose u_2(s)}  +F_K(0)
{v_1(s)\choose v_2(s)}
\en
and similarly for the strange form factors.
In refs.\cite{dgl}\cite{L6} the normalization at the origin was taken 
from CHPT at $O(p^2)$. Here, we wish to investigate the chiral
corrections to these results, so
we need to go to the next chiral order. 

\subsection{Evolution of $F_\pi$}

In the $SU(2)$ chiral limit,
the strange form factor of the pion vanishes at the origin $G_\pi(0)=0$, 
the LEC $L_4$ is related to the first derivative at the origin of  
$G_\pi$. It is convenient to consider the proportional  quantity
\be\lbl{dfexp}
d^{exp}_F=\sqrt{2\over3}m_sB_0\, G'_\pi(0) 
\equiv \sqrt{2\over3}m_sB_0\,G_K(0)v'_1(0)\ .
\en
Here, $v'_1(0)$ can be determined from experiment by the procedure outlined
above, and one finds,
\be
v'_1(0)\simeq 0.27\ {\rm GeV^{-2}}\ ,
\en
using the T-matrix parametrisation of ref.\cite{kll} (using that
of ref.\cite{au} one would find $v'_1(0)\simeq 0.31$ GeV$^{-2}$).
We next need to express $m_sB_0$ and $G_K(0)$ in eq.\rf{dfexp}. At leading
chiral order, one has  $m_sB_0=m^2_K-m^2_\pi/2$ and 
$G_K(0)=\sqrt2$. 
Including
$O(m_s)$ corrections to this result, one obtains $d^{exp}_F$ in the form,
\be
d^{exp}_F={2\over\sqrt3}\left[m^2_K-m^2_\pi/2 
+ {m^4_K\over F^2_\pi}\left(
-8S_8-16S_6+{1\over36\pi^2}(L_\eta+1)\right)\right]\,v'_1(0),
\en
with
\be
S_8=-2L^r_8(\mu)+L^r_5(\mu),\ 
S_6=-2L^r_6(\mu)+L^r_4(\mu),\ L_\eta=\log{m^2_\eta\over\mu^2}\ .
\en
Corrections of order $m_\pi^2 m^2_K$ 
or  $m_\pi^4$ have been neglected here and CHPT at $O(p^4)$
has been used to obtain $G_K(0)$ and to express $m_sB_0$ in terms of 
the physical Kaon mass.
We observe that low-energy constants appear 
now  but it its consistent to use
their values obtained at $O(p^4)$, i.e. for $L_5$, $L_8$ the
values given in ref.\cite{gl85} and for $L_4$, $L_6$ those
obtained in ref.\cite{L6}. We can then match $d^{exp}_F$ 
with its expression
as a   chiral expansion. The latter is obtained from the 
relation\cite{dgl}
\be
d_F={m_s\over F_\pi}{dF_\pi\over dm_s}
\en
which holds in the chiral $SU(2)$ limit, $m_u=m_d=0$, and one can use the
chiral expansion of $F_\pi$ which has been determined up to 
$O(p^6)$ \cite{kambor}\cite{amoros}. One obtains
\be\lbl{dfchpt}
d_F={m_sB_0\over F^2_0} d^{(4)} 
+{m^2_sB^2_0\over F^4_0} d^{(6)}
+O(m_s^3)\ ,
\en
with
\be
d^{(4)}=8L^r_4-{1\over32\pi^2}(L_K+1),\quad L_K=\log{m_sB_0\over\mu^2},
\en
and
\bea
&d^{(6)}=&64C^r_{16}-128 (L^r_4)^2-{17\over3072\pi^4}L^2_K
-{5\over1152\pi^4}L_KL_\eta\nonumber\\
&&+{L_K\over\pi^2}  \left( {73\over9216\pi^2}+4L^r_1+L^r_2+{5\over4}L^r_3
+{1\over2}L^r_5-2L^r_6-L^r_8\right)\nonumber\\
&&+{L_\eta\over\pi^2}\left(-{1\over768\pi^2}+{16\over9}L^r_1+
{4\over9}L^r_2+{4\over9}L^r_3-{8\over9}L^r_4\right)\\
&&+{1\over\pi^2}\left(
{26\over9}L^r_1+{35\over54}L^r_3+{5\over9}L^r_4+{3\over4}L^r_5
-3L^r_6-{3\over2}L^r_8-{0.0022\over\pi^2}\right)\nonumber\ .
\ena
In this formula, the finite contribution from the so-called sunset
diagrams has been evaluated numerically. Equating
\rf{dfexp} and \rf{dfchpt} at the leading, linear
order in $m_s$, gives the $O(p^4)$ determination of $L_4$. 
From the numerical results
of refs.\cite{dgl}\cite{L6} one obtains
\be\lbl{L4op4}
L^r_4(m_\eta)\simeq 0.6\,10^{-3}\quad [ {\rm order}\ p^4 ]\nonumber\ .
\en
If we include the corrections quadratic in $m_s$ now, a new $O(p^6)$ 
coupling constant appears, labelled $C_{16}$ in ref.\cite{bce2}, 
so strictly speaking, we have one equation
and two unknowns.
A similar situation will prevail in the determination
of $L_6$ to be discussed below. We expect, however, the chiral expansion
to be meaningful and the $O(p^6)$ part of the expansion to be smaller
than the  $O(p^4)$ one. If we assume a given ratio for these two parts we can
determine $L_4$ and $C_{16}$ separately. 
Once these two couplings are known we can, furthermore, using 
the CHPT expressions of ref.\cite{kambor}\cite{amoros}, determine
how $F_\pi$ varies in going from an $SU(2)$ chiral limit to an  $SU(3)$
one. 
Some results are collected
in table 1, where we have varied  the $O(p^6)$ over
$O(p^4)$ ratio between 10\% and 100\%.

This exercise seems to indicate that the determination of $L_4$ is rather
stable and rather close 
to its determination at leading order,
and provides an estimate for the size of the $O(p^6)$
coupling constant $ C^r_{16}$. A priori, however, one cannot completely 
exclude a different solution, with $L_4(m_\eta)=0$, for instance. 
In this case, all the OZI violation would be concentrated in the $O(p^6)$
parameter, which seems somewhat unplausible. Taking into account the
experimental uncertainties in the T-matrix, the uncertainty from the 
energy region above 1 GeV (which were estimated 
in ref.\cite{L6}) and the discussion above of the $O(m_s)$
corrections, we find that the error on $L_4$ should be of the order
of 30\%, i.e.,
\be
L_4(m_\eta)=(0.6 \pm 0.2)\,10^{-3}.
\en 

\TABLE[!ht]{
\centering
\begin{tabular}{|c|c|c|c|}\hline
$O(p^6)/O(p^4)$ & $10^3\,L_4$ & $10^5\,C_{16}$ &
$F/F_0-1$\\ \hline
0.10 & 0.58\ [0.31]  & 0.36\ [0.24] & 0.18+0.005  \\
0.30 & 0.53\ [0.26]  & 0.39\ [0.26] & 0.17+0.011\\
0.50 & 0.49\ [0.22]  & 0.41\ [0.28] & 0.16+0.016\\
1    & 0.42\ [0.15]  & 0.44\ [0.32] & 0.14+0.024\\ \hline
\end{tabular}
\caption{Assuming a given ratio of the $O(p^6)$ to the $O(p^4)$
contributions in the chiral expansion of $d_F$, the values of the
LEC's $L^r_4(\mu)$ and $C^r_{16}(\mu)$ are computed
for $\mu=m_\eta$ (in brackets, $\mu=m_\rho$). We also display
the successive contributions to the ratio of $F\equiv (F_\pi)_{SU(2)}$ over
$F_0\equiv (F_\pi)_{SU(3)}$}
\label{Table1}}

\subsection{Evolution of $\langle\bar u u\rangle$}

Let us now discuss the coupling constant $L_6$. 
Using the same scalar form factors, it was remarked in ref.\cite{L6} that one
can derive an estimate of $L_6$. The idea is to consider
the two-point correlation function,
\be\lbl{Pi6def}
\Pi_6(p^2)={i\over B_0^2}\int d^4x\,{\rm e}^{ipx}\langle 0\vert T
(\bar u u(x)+\bar d d(x))\bar s s(0)\vert 0\rangle_{conn.}\ .
\en
In an energy domain extending roughly up to the $\eta\eta$ threshold, the
spectral function is given in terms of the scalar form factors
of the pion and the Kaon,
\bea
&Im \Pi_6(s)=&\sqrt{s-4m^2_\pi\over s} F_\pi(s) G_\pi^*(s)
\theta(s-4m^2_\pi)\nonumber\\
&&+\sqrt{s-4m^2_K\over s}F_K(s) G_K^*(s)\theta(s-4m^2_K)\ .
\ena
Essentially, a usual assumption is made there, that the $4\pi$ or $6\pi$
contributions are negligibly small in this energy 
region (see, e.g.\cite{au}). In the $SU(2)$ chiral limit, $m_u=m_d=0$, a
super-convergence relation holds,
\be\lbl{supconv}
\int_0^\infty Im \Pi_6(s)\,ds=0\ .
\en
Similar 
chiral sum rules are known to be saturated to a fairly good approximation
in terms of a few low-lying resonances (see, e.g.\cite{dono}). In the 
energy region below 1 GeV, inserting the form factors constructed as discussed
above, the contribution to the integral is found to be positive, and dominated
by the $f_0(980)$ resonance. A plausible assumption, then, is that the
higher energy contribution to the sum rule \rf{supconv} will be saturated, at
least approximately, by the next prominent scalar resonance, the $f_0(1500)$,
which we expect to make a negative contribution. We observe that the signs
of the contributions of the $f_0(980)$ and the $f_0(1500)$ conforms 
with the 
recent assignments of Minkowski and Ochs\cite{mink} as 
(essentially) $SU_F(3)$ singlet
and octet respectively. 
Finally, one can calculate the value of the correlation function
at zero energy, $\Pi_6(0)$, making use of the scalar form factors below
1 GeV, and the super-convergence relation \rf{supconv} as,
\be
\Pi_6(0)\simeq{1\over\pi}\left(\int_0^1{Im\Pi_6(s)\over s}\,ds-{1\over 
s_0}
\int_0^1 Im\Pi_6(s)\,ds\right)\ ,
\en
with $s_0\simeq1.5^2$ GeV$^2$. This gives us an ``experimental'' value of
$\Pi_6(0)$, the difference with ref.\cite{L6} is that we now wish to take 
into account chiral corrections in the normalization of the form factors. 
Below the $K\bar K$ threshold, we can express the spectral function in terms
of the basis solutions to the Muskhelishvili-Omn\`es equations $u_i(s)$, 
$v_i(s)$ as,
\bea
&Im\Pi_6(s)=\sqrt{s-4m^2_\pi\over s}\big[ &F_\pi(0)G_K(0) u_1(s)v^*_1(s)
\nonumber\\
&&+F_K(0)G_K(0) v_1(s)v^*_1(s)\big]\ .
\ena
Including $O(m_s)$ chiral corrections in the normalizing
factors, we now have
\bea\lbl{norm6}
&&F_\pi(0)G_K(0)=2\sqrt3\left(1 +{m^2_K\over F^2_\pi}
\left[-16 S_8-48 S_6+ {1\over24\pi^2}L_\eta +{1\over36\pi^2}\right]\right)
\qquad
\nonumber\\
&&F_K(0)  G_K(0)=2       \left(1 +{m^2_K\over F^2_\pi}
\left[-32 S_8-80 S_6+ {7\over72\pi^2}L_\eta +{1\over24\pi^2}\right]\right),
\ena
where $S_8=-2L^r_8+L^r_5$, $S_6=-2L^r_6+L^r_4$ and we may use 
in this part the
values of the LEC's as determined at $O(p^4)$.

We then use the experimental result for $\Pi_6(0)$ in conjunction 
with its chiral expansion. 
The computation of $\Pi_6(s)$ to $O(p^6)$
is exposed in the appendix. Concerning $\Pi_6(0)$, the chiral
expansion goes as follows
\be\lbl{L6chpt}
\Pi_6(0)=\Pi^{(4)}+{m_sB_0\over  F^2_0} \Pi^{(6)},
\en
with
\be
\Pi^{(4)}=
64 L^r_6-{1\over16\pi^2}\left({22\over9}(L_K+1)+{4\over9}L_{43}\right)
\ ,
\en
and
\be
L_K=\log{m_sB_0\over\mu^2},\ \ L_{43}=\log{4\over3}\ .
\en
This expression generates the $O(p^4)$ determination\cite{L6} of $L_6$,
\be\lbl{L6op4}
L^r_6(m_\eta)\simeq 0.5\,10^{-3}\quad [ {\rm order}\ p^4 ].
\en
The expression of the $O(p^6)$ part in \rf{L6chpt} is,
\bea\lbl{pi66}
&\Pi^{(6)}=&256 (C^r_{20}+3 C^r_{21})-{1\over72\pi^4}L^2_K
-{1\over96\pi^4}L_KL_{43}+{1\over288\pi^4}L^2_{43}\nonumber\\
&&+{L_K\over\pi^2}   \left({64\over3}S_6 +{70\over9}S_8+{16\over9}S_7
-{55\over2592\pi^2}\right)\nonumber\\
&&+{L_{43}\over\pi^2}\left({16\over3}S_6 +{16\over9}S_8+{16\over9}S_7
-{47\over5184\pi^2}\right)\\
&&+{1\over\pi^2}\left({140\over9}S_6+{175\over27}S_8+{8\over27}S_7
-{19\over5184\pi^2}\right)
\nonumber\ .
\ena
with
\be
S_7=3L_7+L^r_8. 
\en
In these formulas, we have set $m_u=m_d=0$.
These expressions exhibit the explicit
$m_s$ dependence (which will prove usefull below as we plan to integrate 
over $m_s$). For numerical application, we may use $m_sB_0=m^2_K-m^2_\pi/2$
in $\Pi^{(6)}$ while in $\Pi^{(4)}$ we have to use the $O(p^4)$
expansions of $m^2_K$ and $m^2_\pi$\cite{gl85}.
Equating the chiral expansion of $\Pi_6(0)$ with its sum  rule
evaluation gives us a linear relation involving the $O(p^4)$ LEC 
$L_6$ and the $O(p^6)$ combination $C_{20}+3C_{21}$. As before, we must
make a hypothesis concerning the convergence of the 
chiral expansion if we want
to separately estimate these two contributions. 

Another interesting
information contained in the correlation function $\Pi_6$
concerns the variation of
the quark condensate in the chiral $SU(2)$ limit as a function of
the strange quark mass:
\be
{d\langle \bar u u\rangle\over dm_s}=-{B^2_0\over 2}\Pi_6(0)\ .
\en
Using the chiral expansion of $\Pi_6(0)$ we can integrate this equation
in the variable $m_s$ from its physical value down to $m_s=0$. 
In this way we obtain
an estimate for how the condensate varies from a chiral $SU(2)$ limit
to a chiral $SU(3)$ limit. This variation is obtained as an expansion
in powers of the physical strange quark mass (or, alternatively, 
in powers of the Kaon mass),
\be\lbl{ratexp}
\langle\bar u u\rangle_{SU(2)}=
\langle\bar u u\rangle_{SU(3)}\left(1+{m_sB_0\over F^2_0} R^{(4)} +
{m^2_sB^2_0\over F^4_0} R^{(6)}\right)\ ,
\en
with
\be\lbl{ratexp1}
R^{(4)}=32L^r_6-{11\over144\pi^2}L_K-{1\over72\pi^2}L_{43} ,
\en
and
\bea\lbl{ratexp2}
&R^{(6)}=& 64(C^r_{20}+3C^r_{21})-{1\over288\pi^4}L^2_K
-{1\over384\pi^4}L_KL_{43}+{1\over1152\pi^4}L^2_{43}\nonumber\\
&&+{L_K\over\pi^2}\left({16\over3}S_6+{4\over9}S_7+{35\over18}S_8-
{19\over10368\pi^2}    \right)\nonumber\\
&&+{L_{43}\over\pi^2}\left({4\over3}S_6+{4\over9}S_7+{4\over9}S_8
-{5\over5184\pi^2} \right)\\
&&+{1\over\pi^2}\left({11\over9}S_6-{4\over27}S_7+{35\over54}S_8\right)
\nonumber\ .
\ena
Using the relation
\be
-2\langle\bar u u\rangle_{SU(2)}={d\over dm}\left(m^2_\pi 
F^2_\pi\right)_{ m=0},\quad  m={1\over2}(m_u+m_d),
\en
one sees that the expansion of the 
condensate can be rederived from the expansion of the 
product $m^2_\pi F^2_\pi$
in powers of $m_s$. We have verified, using the $O(p^6)$ expansion
\cite{kambor}\cite{amoros} of $m_\pi$, $F_\pi$ that the formulas \rf{ratexp},
\rf{ratexp1}, \rf{ratexp2} are exactly reproduced. 

Let us now discuss some numerical results. At first, it is instructive to
compare the value obtained for $\Pi_6(0)$ using the normalization of
the form factors at $O(p^2)$ and the normalization at $O(p^4)$,
\be\lbl{pi6exp}
\Pi_6(0)\simeq0.022\ [O(p^2)\ {\rm norm.} ],\quad
\Pi_6(0)\simeq0.043\ [O(p^4)\ {\rm norm.} ].
\en  
We see that including $O(m_s)$ corrections in the normalizations 
brings a rather substantial change in the result by approximately a factor of 
two. This can be traced to the large numbers appearing in front of the
combination $S_6=-2L_6+L_4$ in the normalization factors \rf{norm6}. 
Evidently, the difference  $-2L_6+L_4$ is very sensitive to a small variation
of $L_6$ or $L_4$. Taking $L_6$ slightly smaller than the central value
\rf{L6op4} would decrease significantly the modification in 
$\Pi_6(0)$ without essentially altering the other results. 
Upon considering the chiral expansion
of $\Pi_6(0)$ now, this result suggests that the $O(p^6)$ contribution
could be substantial and it is likely to be positive. We have collected
some results in table 2 assuming that the ratio of the 
$O(p^6)$ to the $O(p^4)$ contributions is 
positive and ranges from 10\% to 100\%.
In this range, we find that the rate of convergence for the ratio of quark
condensates seems reasonable. 

\TABLE[!ht]{
\centering
\begin{tabular}{|c|c|c|c|}\hline
$O(p^6)/O(p^4)$ & $10^3\,L_6$ & $10^5\,(C_{20}+3C_{21})$ &
$\langle\bar u u\rangle_{SU(2)}/\langle\bar u u\rangle_{SU(3)}-1$\\ \hline
0.10 & 0.73\ [0.56]& 0.29\ [-0.02]& 0.85  + 0.14 \\
0.30 & 0.64\ [0.47]& 0.37\ [0.06] & 0.76  + 0.18 \\
0.50 & 0.57\ [0.41]& 0.43\ [0.12] & 0.69  + 0.21 \\
1    & 0.46\ [0.30]& 0.52\ [0.21] & 0.58  + 0.26 \\ \hline
\end{tabular}
\caption{Assuming a given ratio of the $O(p^6)$ to the $O(p^4)$
contributions in the chiral expansion of $\Pi_6(0)$ the values of the
LEC's $L^r_6(m_\eta)$ and 
the combination $C^r_{20}(\mu)+3C^r_{21}(\mu)$ are computed
for $\mu=m_\eta$ (in brackets, $\mu=m_\rho$).
We also display the successive terms in the expansion of 
$\langle\bar u u\rangle_{SU(2)}/\langle\bar u u\rangle_{SU(3)}-1$.} }

Both $F_\pi$ and $\langle\bar u u\rangle$
show a tendency towards chiral restoration in going from chiral $SU(2)$
to chiral $SU(3)$. This tendency seems much stronger 
for $\langle\bar u u\rangle$. This is in agreement with the arguments of
ref.\cite{descotes} based on the spectrum of the Dirac operator.
One must however bear in mind that its dimensionality is different
from $F_\pi$: if we had considered  $\langle\bar u u\rangle^{1/3}$, we 
would have found a smaller variation. This behaviour could perhaps suggest
the possibility, at larger $N_F$, of a phase with vanishing quark
condensate, yet with chiral symmetry still spontaneously broken\cite{stern}.
Kogan et al.\cite{kogan} have discussed how this could result from
a discrete $Z_2$ axial subgroup remaining unbroken, but argue 
against this possibility in QCD. Concerning the error in this evaluation 
of $L_6$, finally, it is expected to be somewhat larger than the error
on $L_4$ because of a greater sensitivity to the energy region 
above 1 GeV. The discussion above also suggests that one should have
$L_6(m_\eta)< L_4(m_\eta)$ otherwise the $O(m_s)$ corrections could 
become out of control. Keeping this mind, we find that $L_6$ should lie in the
following range,
\be
L_6(m_\eta)=(0.5 \pm 0.3)\, 10^{-3}.
\en

\section{The LEC's in a linear sigma-model}

\subsection{General scalar meson lagrangian}

Here, we adopt a simple resonance saturation
point of view for estimating the low-energy 
coupling constants, which was discussed in detail 
in ref.\cite{egpr}. It
consists in making a tree level calculation starting from a lagrangian
for the resonances. We will consider the scalar resonances here. 
It is most convenient to start from a representation of the resonances 
in which they transform under a non-linear 
representation of the chiral group\cite{cwz}. 
A detailed discussion can be found in ref.\cite{egpr}, we adopt
essentially the same notations here. 
If one 
is interested in the $O(p^4)$ LEC's then one needs  consider only those 
terms which are quadratic in the resonance fields, 
\be
{\cal L}_{SS}={1\over2}\d_\mu S_0\d^\mu S_0 +{1\over2}\langle
\D_\mu S \D^\mu S\rangle 
-{1\over2} M^2_0 S_0^2 -{1\over2} M^2_8 \langle S^2\rangle, 
\en
or containing one 
resonance field and one scalar source, or one scalar field and
two chiral fields $u_\mu$, i.e.
\be\lbl{LS2}
{\cal L}_{S\chi}=
        c_d \langle Su_\mu u^\mu\rangle 
+\tilde c_d S_0 \langle u_\mu u_\mu\rangle
+       c_m\langle  S\chi_+\rangle 
+\tilde c_m S_0 \langle \chi_+\rangle ,
\en
where $S_0$ is the singlet scalar, with chiral limit mass $M_0$, $S$ is 
a traceless
matrix encoding the scalar octet, which has a common mass 
$M_8$ in the chiral
limit. We use the same conventions otherwise as ref.\cite{egpr}. This 
lagrangian yields the resonance saturation estimates for the following 
low-energy constants,
\bea\lbl{op4}
&&L_4=-{c_d c_m\over3 M^2_8}+{\tilde c_d\tilde c_m\over M^2_0},\quad
L_5= {c_d c_m\over  M^2_8},\nonumber\\
&&L_6=-{c^2_m\over6M^2_8}   +{\tilde c^2_m\over2M^2_0},\quad
L_8= {c^2_m\over 2M^2_8}\ . 
\ena
In addition, the scalar resonances make  contributions to $L_1$, $L_3$,
\be 
L_1=-{c_d^2\over6M^2_8}+{\tilde c_d^2\over2M^2_0},\quad
L_3= {c_d^2\over2M^2_8},
\en
which receive other  important contributions from the vector
and axial-vector resonance sector. 
These tree-level estimates yield coupling constants 
which are scale independent\footnote{If one starts from the renormalizable
sigma-model it is possible, in principle, 
to compute the one-loop effective action and then
generate the $O(p^4)$ 
LEC's with the correct scale dependence. This was done for
the SU(2) sigma-model in ref.\cite{gl84}}
: one usually assumes that they should represent
meaningful estimates of the $L^r_i(\mu)$ for values of $\mu$ for which 
chiral logarithms are numerically small, that is of the order of $\mu=0.5$ 
to $\mu=1$ GeV. In ref.\cite{egpr} the OZI rule was assumed to apply, 
implying $L_4=L_6=0$. The formulas \rf{op4} provide one relation
between the values of $L_5$, $L_8$ and the experimental value of the 
decay width $a_0\to\pi\eta$. Values  of the couplings $c_d$ and $c_m$ were
extracted,
\be\lbl{cdnum}
\vert c_d\vert\simeq 32\ {\rm MeV}\quad
\vert c_m\vert\simeq 42\ {\rm MeV}\ .
\en
Below, we will investigate the more detailed predictions that one can 
make if one assumes the validity of a renormalizable, sigma-model type
interaction lagrangian. Before we discuss this, let us consider the 
generalisation of the considerations above to  LEC's of chiral order
$O(p^6)$. 
We will illustrate the complications which arise at this order by considering
only the lagrangian terms which are cubic in the scalar source $\chi_+$. 
These terms involve the three
coupling constants\cite{bce2} $C_{19}$, $C_{20}$, $C_{21}$,
\be
{\cal L}^{(6)}=...+C_{19}\langle \chi_+^3\rangle
+C_{20}\langle \chi_+^2\rangle\langle \chi_+\rangle
+C_{21}\langle \chi_+\rangle^3+...
\en

Firstly, one needs to consider interaction terms which are cubic
in the resonance fields,
\be\lbl{SSS}
{\cal L}_{SSS}= a S_0^3 +b S_0 \langle S^2 \rangle  +c \langle  S^3\rangle ,
\en
then terms which are quadratic in the resonance fields and linear in the
scalar source,
\bea\lbl{SSchi}
&&{\cal L}_{SS\chi}=\emo \langle S^2\rangle \langle \chi_+\rangle +\fmo S_0 \langle S \chi_+\rangle \nonumber\\
&&\quad +\dmo \langle  S^2\chi_+\rangle +\ddmo S^2_0 \langle \chi_+\rangle,  
\ena
and, finally, terms linear in the scalar field and quadratic in the scalar
source,
\bea\lbl{Schi2}
&&{\cal L}_{S\chi\chi}=a' S_0 \langle \chi_+\rangle ^2 +b' S_0 \langle \chi^2_+\rangle +
c'\langle S\chi_+\rangle \langle \chi_+\rangle \nonumber\\
&&\quad +d'\langle S\chi^2_+\rangle .
\ena
This already brings in  a  large number
of couplings and it would not be possible to make definite predictions
without additional assumptions. The terms in eqs.\rf{SSS},\rf{SSchi}
and \rf{Schi2} generate chiral $O(p^6)$ coupling 
constants as well as 
chiral symmetry breaking for the scalar meson masses. 
It is convenient, both for generating the $O(p^4)$ and $O(p^6)$
LEC's and for expressing  the scalar meson masses to make the
following field redefinitions,
\bea\lbl{redef}
&&S_0\to S_0+{a\over M^2_0}S^2_0 +{b\over M^2_0}\langle S^2\rangle 
+{\tilde c_m\over M^2_0} \langle \chi_+\rangle \nonumber\\
&&S\to S +{c\over M^2_8}\left( S^2-{1\over3}\langle S^2\rangle I_d\right)
+{c_m\over M^2_8}\left(\chi_+-{1\over3}\langle \chi_+\rangle I_d\right)
\ena
The redefined fields $S_0$ and $S$ have, by construction, no trilinear
couplings and no minimal one-resonance to one-source coupling.
The couplings of two resonances to one scalar source, eq.\rf{SSchi}, get
redefined to
\bea\lbl{2r1s}
&&\dm= \dmo +{3 c_m c\over M^2_8},\quad
\ddm=\ddmo+{3 \tilde c_m a\over M^2_0},\nonumber\\
&&\em= \emo+{\tilde c_m b\over M^2_0}-{c_m c\over M^2_8},\quad
\fm = \fmo+{2 c_m b\over M^2_8}\ .
\ena
The scalar meson masses, to linear order 
in the quark masses, can be straightforwardly expressed in terms of 
these redefined parameters. The masses of the isospin $I=1$ and
$I=1/2$ mesons read,
\bea\lbl{mm}
&&M^2_{a_0}=M^2_8-4\em(2m^2_K+m^2_\pi)-4\dm m^2_\pi\nonumber\\
&&M^2_{\kappa_0}=M^2_{a_0}-4\dm (m^2_K-m^2_\pi)\ ,
\ena
while the isospin zero meson masses diagonalise 
the singlet-octet mass matrix which gets parametrised as follows
\bea\lbl{mmatrix}
&&{\cal M}_{11}=M^2_0 -4\ddm(2m^2_K+m^2_\pi)\nonumber\\
&&{\cal M}_{12}=       4\fm\sqrt{2\over3} (m^2_K-m^2_\pi) \nonumber\\
&&{\cal M}_{22}=M^2_{a_0}-{16\over3}\dm (m^2_K-m^2_\pi)\ .
\ena
In deriving these expressions for the masses, we have ignored possible 
terms in the scalar lagrangian of the form 
$\partial_\mu S_0\partial^\mu S_0\langle \chi_+\rangle $,  
$\langle \nabla_\mu S\nabla^\mu S\chi_+\rangle $, 
anticipating on the fact that such terms will not
appear in the model to be discussed below.

Concerning the LEC's $C_{19}$, $C_{20}$,  $C_{21}$, 
the following expressions are obtained,
\bea\lbl{lec6}
&&F^{-2}_0C_{19}={c^2_m\over M^4_8}\left(\dmo +{c\, c_m\over M^2_8}\right)
+d'{c_m\over M^2_8}\nonumber\\
&&F^{-2}_0C_{20}=
{c^2_m\over M^4_8}\left(-{2\over3}\dmo+\emo-{c\, c_m\over M^2_8}\right)
+{c_m\tilde c_m\over M^2_0M^2_8}\left(\fmo+ {b\, c_m\over M^2_8}\right)
\nonumber\\
&&\quad +(c'-{1\over3}d'){c_m\over M^2_8}+b'{\tilde c_m\over M^2_0}\nonumber\\
&&F^{-2}_0(C_{20}+3C_{21})
=-{c^2_m\over 3M^4_8}\left(\dmo +{c\, c_m\over M^2_8}\right)
+{3\tilde c_m^2\over M^4_0}\left(\ddmo+ {a\,\tilde c_m\over M^2_0}\right)
\nonumber\\
&&\quad +(3a'+b'){\tilde c_m\over M^2_0}-d'{c_m\over3M^2_8}\ .
\ena

\subsection{Parameters of the linear sigma-model}

In the $SU(3)$ linear sigma-model, one first encodes the scalar nonet and the 
pseudo-scalar nonet into a complex $3\times3$ matrix $\Sigma$, which, under
a chiral transformation transforms as,
\be
\Sigma\to g_R \Sigma g^\dagger_L\ ,
\en
and the 
lagrangian is assumed to be renormalizable. In the chiral limit, the most
general renormalizable lagrangian, invariant under the chiral 
group (except for the $U_A(1)$ subgroup) contains four parameters,
\bea\lbl{lsm0}
&&{\cal L}={1\over2}\langle \d_\mu\Sigma\, \d^\mu\Sigma^\dagger\rangle 
-{1\over2}\mu^2\langle \Sigma\Sigma^\dagger\rangle 
-\lambda \langle (\Sigma\Sigma^\dagger)^2\rangle \nonumber\\
&&-\lambda'\langle \Sigma\Sigma^\dagger\rangle ^2
+\beta\,( {\rm det}(\Sigma)+{\rm det}(\Sigma^\dagger))\ .
\ena
This lagrangian was reconsidered recently 
by several authors
\cite{randrup}\cite{ishida}\cite{napsuciale}\cite{tornqv} 
and we use the same notations as in ref.\cite{tornqv}.  
Suitable choices of the parameters $\mu^2,\ \lambda,\ \lambda',\ \beta$ 
ensure that spontaneous chiral 
symmetry breaking occurs, i.e. that the potential is minimised for a non-zero
value of the vacuum expectation value of the $\Sigma$ matrix, 
\be
\bar \Sigma=v I_d,
\en
(which we take to be diagonal,
assuming that no spontaneous breaking of flavour symmetry occurs). 
Values of $v$ which correspond to
extremums of the potential are solutions of the cubic equation,
\be\lbl{eq1}
v\left((4\lambda+12\lambda')v^2-2\beta v+\mu^2\right)=0,
\en
which has real solutions $v\ne0$ provided the parameters satisfy
\be\lbl{vcond}
\beta^2\ge \mu^2(4\lambda+12\lambda')\ .
\en
Next, 
expanding around the minimum we find that the pseudoscalars are massless,
except for the singlet, $\eta_0$ which mass is proportional to $\beta$,
\be\lbl{eq2}
M^2_{\eta_0}=6\beta v\ .
\en
We can also read off the expressions for the
singlet and octet scalar meson masses in the chiral limit, $M_0$ and $M_8$,
\be\lbl{eq3}
M^2_8= 8\lambda v^2 +4\beta v,\quad M^2_0=8(\lambda+3\lambda')v^2-2\beta v\ .
\en
Stability of the vacuum requires that these  squared masses be positive. 
Using eqs.\rf{eq1}\rf{eq2} and \rf{eq3} we can trade the original four 
parameters of the lagrangian \rf{lsm0} for the more physically
relevant ones $v$, $M_{\eta_0}$, $M_0$ and $M_8$. 

In QCD, chiral symmetry is broken by the light quark masses. We will make
the simplifying assumption, at the level of the sigma-model, to consider
symmetry breaking to linear order in the quark masses but we accept
all the terms which are soft (i.e. of dimensionality strictly smaller
than four). Under these assumptions
the most general symmetry breaking
sector has, again, four parameters,
\bea\lbl{sb}
&&{\cal L}_{SB}=\gamma_0 \langle \Sigma \chi^\dagger\rangle  +
\gamma_1 \langle \Sigma^{-1}\chi\rangle {\rm det}(\Sigma) 
+\gamma_2 \langle \Sigma\Sigma^\dagger\Sigma\chi^\dagger\rangle \nonumber\\
&&\quad +\gamma_3 \langle \Sigma\Sigma^\dagger\rangle \langle 
\Sigma\chi^\dagger\rangle +h.c.\ .
\ena
Usually, only the first term is considered. 
This increased phenomenological flexibility will allow us to obtain better
fits of the scalar nonet masses.
While more 
general, this lagrangian will prove nevertheless to be reasonably
constraining. 
We note that only the first two terms are renormalizable in the strict sense
that the counterterms are exactly of the same form\cite{symanzik}. 
The last two terms
generate counterterms which are of higher order in the scalar source $\chi$. 
Consistent with the assumption of not including further terms
quadratic in the quark mass matrix at this level, we will compute the scalar
meson masses and the vacuum expectation values at linear order as well. We
have checked  this approximation 
by comparing with exactly calculated masses.

The connection with the representation of the scalar fields 
used in sec.3.1 is performed by making a
change of variables\cite{weinberg} 
\be\lbl{wtrans}
\Sigma=\exp\left(i\sqrt{2\over3}{\eta_0\over F_0}\right)\,u H u 
\en
with
\be\lbl{wtrans1}
u=\exp\left(i\sum_1^8 {\lambda_a\pi_a\over2F_0}\right),\quad
H=(v+{S_0\over\sqrt{3}})I_d +S\ .
\en
This change of variable makes sense only if $F_0\ne0$, i.e. 
when chiral symmetry is spontaneously broken.
The kinetic energy part of the linear sigma-model lagrangian gets transformed
into
\bea\lbl{lkin}
&{\cal L}_{kin}=&
{1\over2}\d_\mu S_0\d^\mu S_0 +{1\over2}\langle \D_\mu S \D^\mu S\rangle +
{1\over8}\langle \{u_\mu,H\}^2\rangle 
+{1\over3F^2_0}\d_\mu\eta_0\d^\mu\eta_0\langle H^2\rangle
\nonumber\\
&&-\sqrt{2\over3}{\d_\mu\eta_0\over F_0}\langle H^2 u^\mu    \rangle
\ena 
with
\be
\D_\mu S=\d_\mu S+[\Gamma_\mu,S],\ 
\Gamma_\mu={1\over2}(u^\dagger\d_\mu u +u \d_\mu u^\dagger),\ 
u_\mu     =i(u^\dagger\d_\mu u -u \d_\mu u^\dagger).
\en
Replacing $H$ by its expression above \rf{wtrans1}
one identifies $v$ with the pion decay 
constant in the chiral limit 
$F_0$ and one obtains a prediction for the two couplings $c_d$ and
$\tilde c_d$,
\be\lbl{cd}
v^2={F^2_0\over2},\quad c_d=v,\quad \tilde c_d={v\over\sqrt{3}}\ .
\en
The symmetry breaking lagrangian, expressed in terms of the new fields
reads
\bea
&{\cal L}_{SB}=&\langle
(\gamma_0 H+\gamma_2 H^3+\gamma_3\langle H\rangle^2 H)
(\chi_+\cos\sqrt{2\over3}{\eta_0\over F_0}-i
 \chi_-\sin\sqrt{2\over3}{\eta_0\over F_0})\rangle\nonumber\\
&&+ \gamma_1{\rm det}H\langle H^{-1}
(\chi_+\cos\sqrt{8\over3}{\eta_0\over F_0}+i
 \chi_-\sin\sqrt{8\over3}{\eta_0\over F_0})\rangle
\ena
In this sector, we obtain one relation among the 
four parameters $\gamma_i$ from the requirement that the 
coefficient of the chiral lagrangian term 
$\langle \chi_+\rangle $ be correctly normalized to $F^2_0/4$ 
and we can express two other parameters in terms of the   
meson-source couplings $c_m$ and $\tilde c_m$. This gives the relations
\bea
&&\gamma_0= {3\over4}v+{1\over6} c_m
-{2\over\sqrt{3}}\tilde c_m+\gamma_3 v^2
\nonumber\\
&&\gamma_1 v=-{1\over3}(c_m-\sqrt{3}\tilde c_m)-2\gamma_3v^2
\nonumber\\
&&\gamma_2 v^2=-{1\over4}v+{1\over6}c_m+{1\over\sqrt{3}}\tilde c_m
-2 \gamma_3 v^2\ .
\ena
In this sector, we can take $c_m$, $\tilde c_m$ and $\gamma_3$ as
arbitrary parameters. 

In this model, we can now express the many parameters that appeared
in the general scalar lagrangian discussed above. Firstly, for the terms
coupling one scalar field to two scalar sources we have simply 
$a'=b'=c'=d'=0$. For the trilinear couplings of resonances, 
one obtains the relations
\bea
&&a=-{1\over2\sqrt3 v}\left(M^2_0+{1\over9} M^2_{\eta_0}\right)
\nonumber\\
&&b=-{1\over2\sqrt3 v}\left(M^2_0+2M^2_8-{2\over3} M^2_{\eta_0}\right)
\nonumber\\
&&c=-{1\over2v}\left(M^2_8-{8\over9} M^2_{\eta_0}\right)\ .
\ena
Finally, the parameters $\dm,\ddm,\em,\fm$ 
which control linear 
symmetry breaking of the scalar meson masses (see eqs.\rf{mm},
\rf{mmatrix}) obey the following relations,
\bea\lbl{dmmsl}
&&\dm={4\over3v}\left(\sqrt3 \tilde c_m -c_m +
c_m{M^2_{\eta_0}\over M^2_8}\right)-{3\over4}-8\gamma_3 v\nonumber\\
&&\sqrt3\fm={1\over v}\left(\sqrt3 \tilde c_m-c_m {M^2_0\over M^2_8}\right)
-{9\over8}+{1\over2}\dm\nonumber\\
&&\em={M^2_{\eta_0}-3M^2_8\over9M^2_0 v}
\left(\sqrt3 \tilde c_m-c_m {M^2_0\over M^2_8}\right)-{3\over16}
-{1\over4}\dm\nonumber\\
&&\ddm=-{M^2_{\eta_0}\over18M^2_0 v}
\left(\sqrt3 \tilde c_m-c_m {M^2_0\over M^2_8}\right)-{9\over32}
-{1\over24}\dm\ .
\ena
Making use of the relations \rf{2r1s} we can  express the three $O(p^6)$
LEC's $C_{19}$, $C_{20}$, $C_{21}$ discussed above in terms of the parameters
of the linear sigma-model. We are also interested in the prediction 
for the OZI suppressed $O(p^6)$ LEC $C_{16}$ which participates in the 
flavour variation of the order parameter $F_\pi$ as discussed above. After
a small calculation, the following expression is obtained,
\bea\lbl{C16}
&F_0^{-2} C_{16}=& {1\over54}(2c_m+10\sqrt3\tilde c_m-9v)
\left({1\over M^2_0}-{1\over M^2_8}\right)
\left({2c_m\over M^2_8}+{\sqrt3\tilde c_m\over M^2_0}\right)\ 
\nonumber\\
&&+{\gamma_3 v^2\over9}\left(
{2\sqrt3\tilde c_m\over M^2_0}\left({1\over M^2_0}+{2\over M^2_8}\right)-
{4c_m             \over M^2_8}\left({2\over M^2_0}-{5\over M^2_8}\right)
\right)\\
&&+{M^2_{\eta_0}\over54 M^2_0 M^2_8}
\left({2c_m\over M^2_8}+{\sqrt3\tilde c_m\over M^2_0}\right)
\left(2c_m(1-2{M^2_0\over M^2_8})-\sqrt3 c_m{M^2_8\over M^2_0}\right)\ .
\nonumber
\ena

\subsection{Phenomenological applications}

Altogether, the sigma-model as considered here has six independent parameters
(apart from $F_0$): three chiral limit masses $M_0$, $M_8$ and $M_{\eta_0}$
and three couplings to the scalar source $c_m$, $\tilde c_m$ and $\gamma_3$.
The parameter $M_{\eta_0}$, i.e. the
mass of the $\eta'$ in the chiral limit can be estimated 
to be\cite{egpr},
\be
M_{\eta_0}\simeq 0.8\ {\rm GeV}\ .
\en

Let us now discuss the determination of the remaining five 
parameters. 
A priori, one expects to be able to 
reproduce exactly the four independent masses 
($M_{a_0}$, $M_{\kappa_0}$, $M_{\sigma_0}$, $M_{f_0}$)
in the scalar nonet and, as
an additional constraint a natural choice would be to enforce the
super-convergence relation \rf{supconv} which implies here
$c_m=\sqrt3\tilde c_m$. In practice, however, because of non-linearities,
this choice strongly restricts the possible range of 
masses for the $\kappa_0$.
This is an indication that  this model should be considered as a toy model
rather than a really physically meaningful one.
Instead, we did not try to impose the
super-convergence relation and use a constraint from the pseudo-scalar
sector, the ratio $F_K/F_\pi$ which allows for a fairly broad range 
of values for  $M_{\kappa_0}$.
Because of the specific parametrisation of the mass matrix,
eq.\rf{mmatrix} its entries get evaluated as follows
in terms of physical masses,
\be
{\cal M}_{22}={4M^2_{\kappa_0}-M^2_{a_0}\over3},\quad
{\cal M}_{11}=M^2_{\sigma_0}+M^2_{f_0}-{\cal M}_{22},
\en
and, using the determinant,
\be\lbl{M12}
{\cal M}_{12}=\pm\sqrt{-({\cal M}_{22}-M^2_{\sigma_0})
({\cal M}_{22}-M^2_{f_0})}\ .
\en
We observe that there are two possible sign assignments. Reality of
${\cal M}_{12}$ requires that the following inequality be 
satisfied
\be\lbl{ineq}
{M^2_{a_0}+3M^2_{\sigma_0}\over4}\le M^2_{\kappa_0}\le
{M^2_{a_0}+3M^2_{f_0}\over4}\ ,
\en
As shown in \cite{fariborz2}, this inequality holds in a model 
independent way if symmetry breaking is assumed to be linear.
From the expression of
the mass matrix \rf{mmatrix} it is then easy to see that the 
scalar singlet chiral mass $M^2_0$ satisfies a quadratic equation
\be\lbl{eqM0}
M^4_0-B M^2_0+C M^2_{\eta_0}=0,
\en 
with
\bea
&&B={\cal M}_{11}-{9\over8}(2m^2_K+m^2_\pi)
\left(1-{1\over27}{M^2_{\kappa_0}-M^2_{a_0}\over m^2_K-m^2_\pi}\right)
\nonumber\\
&&C={1\over4}(2m^2_K+m^2_\pi)
\left(1+{1\over9}{M^2_{\kappa_0}-M^2_{a_0}\over m^2_K-m^2_\pi}
+{\sqrt2\over3}{{\cal M}_{12}\over m^2_K-m^2_\pi}\right)\ .
\ena
Real solutions exist provided 
\be\lbl{discr}
B^2-4CM^2_{\eta_0}\ge0 ,
\en
and it must further be checked that at least one solution is positive.
This puts further 
constraints on the allowed values of the scalar meson masses
in this model. Once $M^2_0$ is determined, the other parameters are 
easily evaluated. For instance $M^2_8$ is given by,
\bea
&&M^2_8={1\over 1+{6C\over M^2_0}}\Bigg[
M^2_{a_0}+{2C M^2_{\eta_0}\over M^2_0}-{3\over4}(2m^2_K+m^2_\pi)
\nonumber\\
&&\qquad+{1\over4}(2m^2_K-3m^2_\pi)
{M^2_{\kappa_0}-M^2_{a_0}\over m^2_K-m^2_\pi}\Bigg]\ .
\ena
Next, using as experimental input $M^2_{\kappa_0}-M^2_{a_0}$ 
and ${\cal M}_{12}$ together with eqs.\rf{mm}, 
\rf{mmatrix} and \rf{dmmsl}) gives two linear equations for the three
quantities $c_m/v$, $\tilde c_m/v$ and $\gamma_3 v$. We will use as
an additional constraint, the ratio $F_K/F_\pi$, which determines
$c_m/v$ from the relation,
\be
{c_m\over v}={M^2_8\over 2F_\pi}\,{F_K-F_\pi\over m^2_K-m^2_\pi}\ .
\en
Finally, there remains to determine 
$v=F_0/\sqrt2$ . To linear order in the quark
masses, $v$ is given by the following expression
\be\lbl{v}
v={F_\pi\over\sqrt2}\left[1
-{2\over3}m^2_\pi\left({2c_m/v\over M^2_8}+{\sqrt3\tilde c_m/v\over M^2_0}
\right)
-{4\over3}m^2_K  \left({-c_m/v\over M^2_8}+{\sqrt3\tilde c_m/v\over M^2_0}
\right)\right]\ .
\en
This expression can be recovered in two different ways: one can either
use CHPT together with the expressions \rf{op4} for $L_4$ and $L_5$ or write
down the equations for the vacuum expectation values in the presence of quark
masses and solve these equations to linear order in the quark masses.

\subsubsection{ Light $\sigma$ meson}

The existence of a very 
broad scalar resonance in $\pi\pi$ scattering
with $M\simeq\Gamma\simeq 0.5-0.6$ GeV 
is by now well established since the work of Basdevant Froggatt 
and Petersen\cite{basdevant}, in which the whole set of constraints from
unitarity, analyticity, and crossing symmetry has been implemented
(see ref.\cite{pennington} for a recent survey and 
the particle data book\cite{pdg} for a complete list of
references). What is unclear is whether this state should be interpreted as
a physical light scalar resonance. 
Recently , Black et al.\cite{fariborz1} have proposed arguments based
on perturbative unitarity favouring the existence of a light
$\sigma$ and also, of a light $\kappa$ meson.
According to the inequality \rf{ineq}, 
if the sigma meson is light then
one must have $M_{\kappa_0}\le M_{a_0}$ unless our
assumption of linearity in the quark masses fails. Recently, T\"ornqvist 
\cite{tornqv} (similar fits were also discussed in ref.\cite{randrup})
has attempted to accommodate in the linear sigma-model 
a scalar meson multiplet
with a light sigma and a heavy $\kappa$. Such fits fail to obtain the
$f_0(980)$ at the correct mass and therefore would not correctly
reproduce the $\pi\pi$ phase-shifts around 1 GeV, even if
one-loop corrections are included\cite{haymaker}.

Let us now assume that a light sigma meson exists, e.g.
$M_{\sigma_0}\simeq0.6$ GeV,  and discuss the consequences.
We take otherwise $M_{a_0}=0.983$ GeV, 
and $M_{f_0}=0.980$ GeV from experiment and $M_{\kappa_0}=0.9$ GeV
as proposed in ref.\cite{fariborz1}. From this input one easily calculates
the parameters $A$ and $B$ in the quadratic equation \rf{eqM0},
\be
A=0.024\ {\rm GeV^2}\quad B=0.042\ {\rm GeV^2}\ .
\en
There results that no real solution to the equation for $M^2_0$ exists
unless $M_{\eta_0}\le0.06$ GeV, which appears as an absurdly small value.
Reality and positivity
of $M^2_0$ (and $M^2_8$) are necessary conditions for the existence
of a stable minimum of the potential, with spontaneous chiral 
symmetry breaking. One must thus conclude that with such input scalar meson
masses $N_F=3$ chiral symmetry is not spontaneously broken in the linear
sigma-model. The same conclusion holds if one picks up larger values for
$M_{\kappa_0}$: using the set of parameters determined in ref.\cite{randrup}
for $M_\sigma=0.4$ and $M_\sigma=0.6$ GeV one verifies that the equation
for the vacuum expectation value $v$ eq.\rf{eq1} has no real solution
$v\ne0$. A better situation prevails if we accept smaller values for 
$M_{\kappa_0}$. If we take, for instance, $M_{\kappa_0}=0.750$ GeV, which
is the smallest value allowed by the inequality\rf{ineq}, then, 
eq.\rf{eqM0} has
real solutions provided $M_{\eta_0}\le 0.62$ GeV, which is still reasonable. 
A problem remains, however, that being close to a situation where chiral
symmetry is unbroken for $N_F=3$, while it is in general broken for 
$N_F=2$, the expansion in the strange quark mass is likely to be 
non-converging. Indeed, if we try to compute $F_0$ with, say, 
$M_{\eta_0}= 0.50$ GeV, using our linear
expansion formula \rf{v}, we find $F_0\simeq 28$ MeV, which is very small
compared to $F_\pi$, suggesting that the expansion in $m_s$ is not 
perturbative in this case.

\subsubsection{ Heavy $\sigma$ meson}

Let us now consider the scenario of a heavy sigma meson, i.e. we identify the
sigma meson with the well established resonance 
$f_0(980)$ and we identify the heavier $I=0$ member of the nonet with 
the $f_0(1500)$ (see \cite{mink}).  
With these assignments, we find that a solution
of eq.\rf{eqM0} with real $M^2_0$ exists in the range
\be
M_{\kappa_0}\le 1.37\ {\rm GeV}\ ,
\en
taking $M_{\eta_0}=0.8$ GeV and the negative sign for ${\cal M}_{12}$
(see \rf{M12}).  This means that we cannot exactly reproduce the experimental
mass of the $K^*_0(1430)$ but we can get reasonably close.
We will eventually allow
$M_{\kappa_0}$ to vary somewhat away from the experimental result. 
Numerical values of the parameters $M_0$, $M_8$, $c_m$, $\tilde c_m$, 
$\gamma_3$ as well as $v$ 
determined for several values of $M_{\kappa_0}$ are
collected in table (3).
\TABLE[!ht]{
\centering
\begin{tabular}{|c||c|c|c|c|c|c|}\hline
$M_{\kappa_0}$ & $M_0$  & $M_8$ & $c_m$ & $\sqrt3\,\tilde c_m$ &
$\gamma_3$     & $\sqrt2\,v$    \\ \hline
1.10 & 1.17 & 0.85 & 0.023 & 0.040 & 0.54 & 0.092\\
1.20 & 1.05 & 0.93 & 0.028 & 0.027 & 0.33 & 0.094\\
1.30 & 0.87 & 0.90 & 0.023 & 0.034 & 1.56 & 0.081\\
1.33 & 0.79 & 0.85 & 0.017 & 0.037 & 2.79 & 0.068\\ 
1.35 & 0.72 & 0.80 & 0.011 & 0.033 & 4.82 & 0.051\\ \hline 
\end{tabular}
\caption{\sl Determination of parameters in the case of a heavy scalar 
nonet $M_{\sigma_0}=0.980$, $M_{f_0}=1.5$  for 
several values of $M_{\kappa_0}$. All entries are in GeV except 
$\gamma_3$ which is in $GeV^{-1}$. }
\label{Table3}}
We note that the situation with a minimal symmetry breaking lagrangian
(i.e. $\gamma_1= \gamma_2=\gamma_3=0$ ) is close to the case when
$M_{\kappa_0}=1.20$ in the table. In this case, one has 
$c_m=\sqrt3\tilde c_m=v/2$. The value of $F_0=\sqrt2 v$ remains 
reasonable close to $F_\pi$ except when $M_{\kappa_0}$ gets 
very near to its upper bound. The coupling $c_m$ comes out smaller
than that determined in ref.\cite{egpr} (see \rf{cdnum})
while the coupling $c_d=v$ comes out larger. The coupling $c_d$
controls the decay width $a_0\to\pi\eta$, which tends to be too large
in this model 
(accepting that the experimental value is $\Gamma\simeq 60$ MeV, which
is subject to some debate\cite{pdg}), 
the result in this respect improves as $M_{\kappa_0}$
gets larger.

Once these couplings are known it is easy to determine the LEC's
with the formulas given above. Numerical values for the
LEC's dominated by the scalar mesons
and not suppressed in the large $N_c$ limit
are shown in table (4). We have also included  $L_7$, which
is assumed to be saturated by the $\eta_0$, computed from the
expression\cite{egpr},
\be
L_7=-{\tilde d_m^2\over2 M^2_{\eta_0}},
\en
where the coupling $\tilde d_m$ of the $\eta_0$ field to the pseudo-scalar
current has the following expression in the present version of the
linear sigma-model,
\be
\tilde d_m=-{F_0\over 2\sqrt6}\left(1+{2\over v}(c_m-\sqrt3\tilde c_m)+
12\gamma_3 v\right)\ .
\en
\TABLE[!bt]{
\centering
\begin{tabular}{|c||c|c|c||c|c|}\hline
$M_{\kappa_0}$ & $10^3\,L_5$    & $10^3\,L_7   $  
               & $10^3\,L_8$ & $10^5\,C_{19}$ & $10^5\,C_{20}$ \\ \hline
1.10 &  2.06 &  -0.22&  0.36&  -0.17&  -0.05   \\
1.20 &  2.15 &  -0.47&  0.45&  -0.34&   0.07  \\
1.30 &  1.61 &  -0.62&  0.32&  -0.45&   0.12  \\ 
1.33 &  1.13 &  -0.51&  0.20&  -0.38&   0.08  \\
1.35 &  0.64 &  -0.31&  0.10&  -0.24&   0.04  \\ \hline
\end{tabular}
\caption{\sl  Numerical values of some $O(p^4)$ and $O(p^6)$
LEC's from the sigma-model with a heavy sigma. }
\label{Table4}}
The fact that $L_5$ does not remain constant when $M_{\kappa_0}$ increases
is a reflection of the effect of non-linearities in $m_s$.
The values of the LEC's may be compared with the ones determined at $O(p^4)$
from experimental data\cite{gl85}
\be
10^3\,L^r_5(m_\eta)= 2.2\pm0.5, \quad
10^3\,L_7        =-0.4\pm0.15,\quad
10^3\,L^r_8(m_\eta)= 1.1\pm0.3\ .
\en
While the prediction for $L_7$ is acceptable, that for $L_8$ is
too small.  This is related to the small size of the
meson to source coupling $c_m$ predicted by the model. 
Still, the order of magnitude is correct except, perhaps, for the
very last line in the table.
The LEC's were recently
re-determined in an $O(p^6)$ analysis of the experimental 
data\cite{amoros1} and this brings the values somewhat 
down in magnitude:
$10^3\,L^r_5(m_\eta)= 1.45\pm0.12,\  
10^3\,L_7= -0.31\pm0.15,\ 
10^3\,L^r_8(m_\eta)=  0.68\pm0.18$. 
A resonance-saturated 
estimate for the $O(p^6)$ coupling $C_{19}$ was provided in 
ref.\cite{amoros}. 
The value that we obtain is significantly smaller, by a factor
$3-4$. One difference in the estimates is that we include the effect
of cubic interaction terms, but the numerical influence of these terms
turns out to be unimportant. The main reason for 
our smaller result is, again, the fact that the coupling $c_m$ is
smaller here.

Let us now return to
the OZI suppressed coupling constants $L_4$ and $L_6$
and their $O(p^6)$ counterparts $C_{16}$ and $C_{20}+3C_{21}$. 
The predictions from the linear sigma-model are collected in 
table (5).
%
\TABLE[!hbt]{
\centering
\begin{tabular}{|c||c|c||c|c|}\hline
$M_{\kappa_0}$ & $10^3\,L_4   $ & $10^3\,L_6             $    
               & $10^5\,C_{16}$ & $10^5\,(C_{20}+3C_{21})$ \\ \hline
1.10 & -0.06 & 0.07 & 0.19 & 0.02\\
1.20 & -0.16 &-0.04 & 0.08 & 0.05\\
1.30 &  0.33 & 0.15 & 0.21 & 0.08\\
1.33 &  0.56 & 0.28 & 0.36 & 0.15\\
1.35 &  0.55 & 0.31 & 0.41 & 0.27\\ \hline
\end{tabular}
\caption{\sl  Numerical values of OZI suppressed $O(p^4)$ and $O(p^6)$
LEC's from the sigma-model with a heavy sigma. }
\label{Table5}}
The values and signs of $L_4$, $L_6$ are seen to depend 
very much on the mass of $\kappa_0$.
They both approximately  vanish when
$M_{\kappa_0}\simeq1.24 $ GeV. This is the point of minimal OZI
violation.
As one pushes the $\kappa_0$ to higher
masses, as required by experiment, then the pattern of OZI violation
is not unlike the one found on the basis of sum rules.
Both $L_4$ and  $L_6$ become
positive, and the orders of magnitudes seem to be in agreement
with the discussion based on sum rules. There is even a 
relatively good agreement
as far as $L_4$ and $C_{16}$ are concerned. 

\section{ Conclusions}

In this paper, we have pursued an investigation on the evolution of 
order parameters of the QCD chiral vacuum as one increases the number
of massless flavours, as can be inferred from experimental data. 
An information on this evolution is contained in
the $O(p^4)$ coupling constants $L_4$ and $L_6$ which control how
$F_\pi$ and $\langle\bar u u\rangle$ respectively vary from the $SU(2)$ chiral
limit with $m_s\simeq 200$ MeV to the $SU(3)$ chiral limit. Estimates
for the couplings $L_4$ and $L_6$ are based on the scalar form factors of
the pion and the Kaon which are reconstructed from experimental 
S-wave scattering data modulo some assumptions. We have investigated the 
$O(m_s)$ corrections to our previous results. These have two origins:
a) the normalization of the form factors at the 
origin must include the $O(p^4)$ contributions 
and b) the chiral expansions 
(for $F_\pi$ and $\Pi_6(0)$ ) must be pursued up to order $p^6$.

These corrections seem not to affect in a significant way the results obtained
previously. In particular there is no sign that these corrections go in the
sense of decreasing the values of $L_4$ and $L_6$ to make them compatible
with the naive large $N_c$ expectation. This conclusion, however, only
holds provided one makes the assumption that the chiral expansion is 
reasonable, in other terms, that the $O(p^6)$ contribution (say in the
expansion  of $\Pi_6(0)$) is not larger than the $O(p^4)$ contribution. 
Modulo this assumption, one also obtains from this analysis an estimate
of the $O(p^6)$ coupling constant $C_{16}$ and the combination 
$C_{20}+3 C_{21}$. 
Influence of the values of $L_4$, $L_6$ on the convergence rate
of the chiral expansion for various observables 
was studied very recently in \cite{amoros2}.
Because of such assumptions which enter into the
calculation, the error bars on our 
coupling constant estimates must be considered as educated guesses. 
Our results are compatible with a tendency
towards chiral  restoration upon increasing $N_F$ from
$N_F=2$ to $N_F=3$ which is significant, particularly so for
the quark condensate (in agreement with the arguments of \cite{descotes}),
for which the decrease by a factor of approximately 
two found in ref.\cite{L6} is confirmed. 
This behaviour of the quark condensate is in qualitative agreement with
the one obtained from the instanton liquid model\cite{schafer}.
It would be also interesting to
compare this result with unquenched lattice simulations with $N_F=3$. 
Unfortunately, such simulations are not available yet, but will exist
in the near future. 

We also discussed some predictions from the linear sigma-model, considered
as a simple toy model. One application is to gauge the possible
importance of self-couplings of resonances (which are present in the model)
in resonance saturation estimates of the $O(p^6)$ LEC's. Furthermore, the
model provides a connection between simply the spectrum of the scalars
and the evolution of  chiral order parameters with $N_F$.
In order to improve the ability of the model
to reproduce experimental masses, the symmetry breaking sector
was made 
more general than usually done, but still linear in the quark mass matrix.
We have considered two possibilities for the light scalar nonet: \\
a)The nonet contains a light $\sigma$ and a light $\kappa$. In this case, we
find that chiral symmetry tends to be restored already at $N_F=3$ or, at
least, for a value of $N_F$ somewhat too small, making the chiral expansion
in $m_s$ unreliable.\\ 
b) The nonet
is composed of the resonances $a_0(980)$, $f_0(980)$,
$K^*_0(1430)$ and $f_0(1500)$ (which are well established). 
In this case, we find that we need only a small
amount of nonlinear effects to exactly reproduce the $K^*_0(1430)$ mass,
and $SU(3)$ chiral symmetry is realised in the Goldstone mode. 
Concerning  $L_4$ and $L_6$, we find that
they become different from zero as one increases the mass of the  
$K^*_0(1430)$ towards its experimental value. The signs and orders of 
magnitude, then, are compatible with the preceding analysis. The same is
true of the related $O(p^6)$ constants.

\acknowledgments{
The author thanks B. Ananthanarayan for drawing his attention to 
ref.\cite{mink} and Jan Stern for useful remarks. 
This work is  supported in part by funds provided by the U.S. Department
of Energy (D.O.E.) under cooperative research agreement \#DF-FC02-94ER40818
and by the EURODAPHNE network EEC contract \#TMR-ERBFMRX-CT980169.}

\appendix
\section{The correlation function $\Pi_6(s)$ to $O(p^6)$}

At chiral order $p^4$, 
a simple calculation gives the correlation function $\Pi_6(s)$,
\be\lbl{oneloop}
\Pi_6(s)=2\bar J_K(s)+{4\over9} \bar J_\eta(s)+64L^r_6 -\sezpi\left[
2L_K + {4\over9} L_\eta +{22\over9}\right],
\en
where $\bar J_P(s)$ is the one-loop function defined as to vanish for $s=0$,
\be
\bar J_P(s)=\sezpi\left( \sigma_P \log{ \sigma_P-1\over\sigma_P+1}+2\right),
\ \sigma_P=\sqrt{1-{4M^2_P\over s}}\ .
\en
In all the formulas  
$M^2_P$ stands not for the physical pseudo-scalar meson 
masses but for their lowest order chiral expansion, 
\be
M^2_\pi\equiv  2mB_0,\quad
M^2_K \equiv  (m+m_s)B_0,\quad 
M^2_\eta\equiv { (2m+4 m_s)B_0\over 3}\ .
\en
The following simplified notation for logarithms was introduced
\be
L_P=\log{M^2_P\over\mu^2}\ .
\en
At order $p^6$, one must compute the one-loop and two-loops
diagrams shown in fig.1 
and add the tree-level contributions from
the $O(p^6)$ chiral lagrangian. The calculation is lengthy but not 
excessively difficult because of the absence of  sunset-type diagrams
in the present case.
The result  may be written as a sum of six terms:
\be
\Delta\Pi_6(s)=\Delta\Pi^a_6(s)+\Delta\Pi^b_6(s)+\Delta\Pi^{resc}_6(s)+
A {s\over F^2_0}+B+C \ .
\en
The first term encodes the $O(p^4)$ corrections to the pseudo-scalar meson 
masses in the one-loop functions,
\be\lbl{mcor1}
\Delta\Pi^a_6(s)=-s\left(
2\,{\Delta M^2_K\over M^2_K} {d\bar J_K(s)\over ds}+
{4\over9}\,  {\Delta M^2_\eta\over M^2_\eta} {d\bar J_\eta(s)\over ds}
\right)\ .
\en
These mass corrections are given by the following expressions\cite{gl85},
\bea
&&\Delta M^2_K=M^2_K\Bigg[ 
{m B_0\over F^2_0}\left(-32 S_6-8 S_8 +{1\over72\pi^2} L_\eta\right)\nonumber\\
&&+{m_s B_0\over F^2_0}\left(-16 S_6-8 S_8 +{1\over36\pi^2} L_\eta 
\right)\Bigg]
\ena
and
\bea
&&\Delta M^2_\eta=M^2_\eta\Bigg[
{m B_0\over F^2_0} \left(-32 S_6-{16\over3} S_8 -{5\over48\pi^2} L_\pi
+{1\over9\pi^2} L_K -{1\over144\pi^2} L_\eta\right)\nonumber\\
&&+{m_s B_0\over F^2_0} \left(-16 S_6 -{32\over3} S_8+{1\over24\pi^2} L_\pi
+{1\over18\pi^2} L_K - {1\over18\pi^2} L_\eta \right)\Bigg]\nonumber\\
&&+{(m_s-m)^2 B^2_0\over F^2_0} \Bigg[
{128\over9} S_7 +{1\over108\pi^2} L_K-{1\over18\pi^2} L_\pi\Bigg]\ .
\ena
In these expressions, $S_6$, $S_7$ and $S_8$ denote the following combinations of
$O(p^4)$ coupling constants,
\be
S_6=-2 L^r_6 +L^r_4,\ 
S_7= 3 L^r_7 +L^r_8,\ 
S_8=-2 L^r_8 +L^r_5\ .
\en

The second corrective term $\Delta\Pi^b_6(s)$ is generated by $O(p^4)$ 
corrections to the scalar form factors and has the following form,
\bea
&&\Delta\Pi^b_6(s)=3\bar J_\pi(s)\Bigg[ {d\Delta M^2_\pi\over dm_sB_0}+
{s\over F^2_0}\left(8L^r_4-\trdpi(L_K+1)\right) \Bigg]\nonumber\\
&&+\bar J_K(s)\Bigg[ 2\,{d\Delta M^2_K\over dm B_0}+
                     2\,{d\Delta M^2_K\over dm_s B_0}+
{s\over F^2_0}\Bigg(16L^r_5+48 L^r_4 
\nonumber\\
&&-\trdpi (3L_\pi+6L_K+3L_\eta+12)\Bigg)
\Bigg]
+{1\over3} \bar J_\eta(s)\Bigg[ 2{d\Delta M^2_\eta\over dm B_0}+
{d\Delta M^2_\eta\over dm_s B_0}
\nonumber\\
&&+{s\over F^2_0}\left( {32\over3}L^r_5+40L^r_4-
{9\over32\pi^2}(L_K+1)\right)\Bigg]\ .
\ena
The derivatives of $\Delta M^2_K$ and $\Delta M^2_\eta$ are easily computed
from the formulas above and for $\Delta M^2_\pi$ one has,
\be
{d\Delta M^2_\pi\over m_sB_0}=
mB_0\left(-32 S_6-{1\over36\pi^2}(L_\eta+1) \right)\ .
\en
The third corrective term  $\Delta\Pi^{resc}_6(s)$ is the rescattering 
contribution,
\bea
&&\Delta\Pi^{resc}_6(s)={3s\over2F^2_0}\bar J_K(s)\left(\bar J_\pi(s)+
\bar J_K(s)+\bar J_\eta(s)\right)\\
&&+{4\over3}\bar J_\eta(s)\left(
   {(m+8m_s)B_0\over 27 F^2_0}\bar J_\eta(s)
   -{(m+m_s)B_0\over F^2_0}\bar J_K(s)
+        {m B_0\over F^2_0}\bar J_\pi(s)\right)\ .
\nonumber
\ena
The last corrective term is a polynomial, linear in $s$. This term
picks up contributions from four  coupling constants of the $O(p^6)$ chiral
lagrangian, 
$C^r_{39}$, $C^r_{20}$, $C^r_{21}$ and $C^r_{94}$ in the notation 
of \cite{bce2}. The part
proportional to $s$  is as follows
\bea
&&A= 64  C^r_{39}+ {1\over\pi^2 }\Bigg[ 
{3\over512\pi^2} L_K(L_\pi+L_K+L_\eta) +
 L_\pi\left(  {3\over512\pi^2}-{3\over 2}L^r_4\right)\nonumber\\
&&+L_K \left ({3\over128\pi^2}        -3L^r_4           -L^r_5\right)
+L_\eta\left( {3\over512\pi^2}-{5\over 6}L^r_4-{2\over9} L^r_5\right)
\nonumber\\
&&+      {9\over512\pi^2}-{16\over3}L^r_4-{11\over9}L^r_5\Bigg]
\ena

The constant terms, finally are

\bea
&&B={mB_0\over F^2_0}\Bigg[32 ( 8C^r_{20}+48 C^r_{21}+C_{94})
-{1\over864\pi^4}L_\eta(-9L_\pi+12L_K-L_\eta)
\nonumber\\
&&+{L_\pi\over\pi^2} \left( 12 S_6 +{1\over96\pi^2}\right)
  +{L_K  \over\pi^2} \left( 20 S_6+6 S_8         -{1\over96\pi^2}\right)
\nonumber\\
&&+{L_\eta\over\pi^2} \left( 4 S_6+{8\over9}S_8-{11\over5184\pi^2}\right)
\nonumber\\  
&&+{1     \over\pi^2}\left({206\over9}S_6-{16\over27}S_7+{124\over27}S_8
                            +{1\over5184\pi^2}\right)\Bigg]
\nonumber\\
&&+{m_s B_0\over F^2_0}\Bigg[ 256( C^r_{20}+3 C^r_{21})
-{1\over288\pi^4}L_\eta(-L_\eta+5 L_K)
\nonumber\\
&&+{L_K\over\pi^2}\left(16 S_6+6 S_8              -{1\over96\pi^2} \right)
+{L_\eta\over9\pi^2} \left(48 S_6 +16 S_7 +16 S_8 -{37\over576\pi^2}\right)
\nonumber\\
&&+{1\over\pi^2}\left({118\over9}S_6+{16\over27}S_7+{140\over27}S_8
-{19\over5184\pi^2}\right)\Bigg],
\ena
and
\be\lbl{mcor2}
C=-\sezpi\left( 2\,{\Delta M^2_K\over M^2_K} +{4\over9}
\,{\Delta M^2_\eta\over M^2_\eta}\right)
\en
The terms proportional to the mass corrections $\Delta M^2_K$ and 
$\Delta M^2_\eta$ (see eq.\rf{mcor1} and \rf{mcor2}) 
may eventually be absorbed into the $O(p^4)$ expression for $\Pi_6(s)$,
which amounts to replace the lowest order expression for the pseudo-scalar 
masses there by their expression up to $O(p^4)$. Formula \rf{pi66} in
the text is obtained from $B+C$, setting $m=0$.

\newpage

\FIGURE{
\epsfig{file=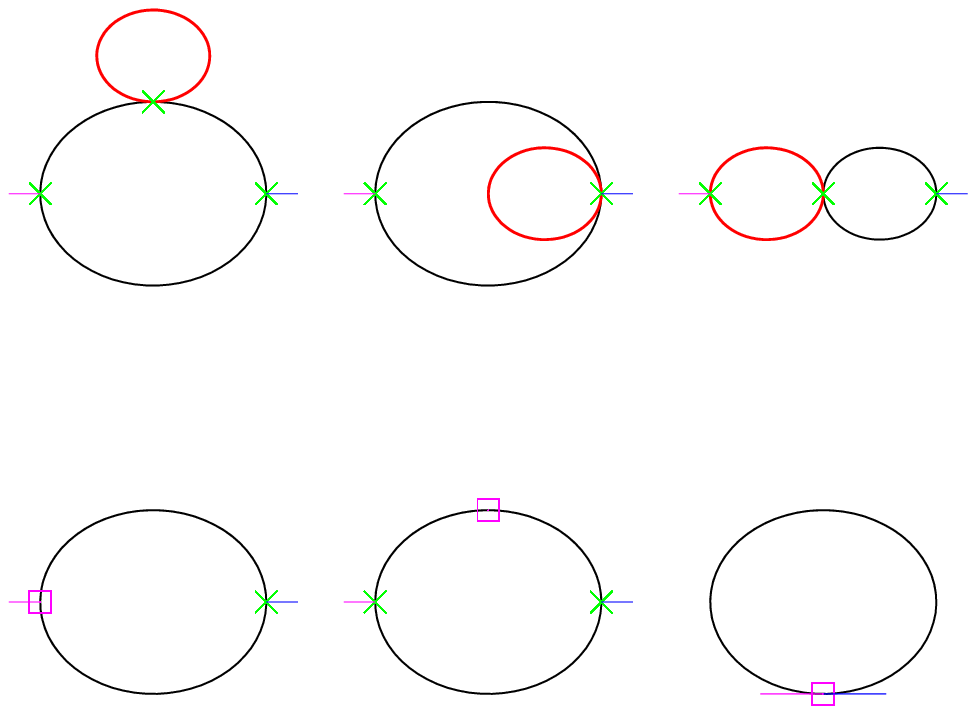,width=\textwidth}
\caption{Set of one and two-loops 
Feynman graphs contributing at chiral order $p^6$ to the
correlation function $\Pi_6(s)$: crosses denote vertices from the $O(p^2)$
chiral lagrangian, boxes denote vertices from the $O(p^4)$ chiral lagrangian. 
Some diagrams obtained by exchanging external lines have not been drawn.}
\label{Fig. 1}}

\end{document}